\documentclass[twocolumn,showpacs,amsmath,amssymb]{revtex4}
\input epsf
\usepackage{appendix}
\usepackage{rotating}
\begin{document}
\title {Excitations of a trapped two component Bose Einstein Condensate}
\author{Christopher Ticknor}
\affiliation{Theoretical Division, Los Alamos National Laboratory, Los Alamos, 
New Mexico 87545, USA}
\affiliation{Kavli Institute for Theoretical Physics, University of California, Santa Barbara, California 93106, USA}
\date{\today}
\begin{abstract}
We present analysis of the excitation spectrum for a 2 component quasi2D 
Bose Einstein Condensate.  We study how excitations change character
across the miscible to immiscible phase transition. We find that 
the bulk excitations are typical of a single-component BEC
with the addition of interface bending excitations. 
We study how these excitations
change as a function of the interaction strength. 
\end{abstract} 
\maketitle
\section{introduction}
Ultracold atoms provide an excellent forum to study complex 
quantum mechanical behavior. An example is the 
superfluid to Mott insulator transition \cite{greiner}, where experimental 
efforts have imaged this as
a quantum phase transition at the single atom level \cite{bakr}. 
A quantum phase transition is a fundamental change in the ground state as a 
parameter is altered, in the superfluid-Mott insulator example the
parameter is lattice depth. 
Near this quantum phase transition the temperature dependence of 
superfluidity and quantum criticality has been studied \cite{trotzky,zhang}.  

Another system in ultracold atoms, which exhibits a quantum phase transition, 
is the 2 component Bose Einstein Condensate (BEC), where by tuning the 
interactions between the components the gas can change from a miscible to an 
immiscible phase, as has been experimentally demonstrated in Rb \cite{papp}.
Theoretical studies showed that if $g_{11}g_{22}-g_{12}^2$ 
is greater (less) than zero, the gas is miscible (immiscible) \cite{pu,eddy}.
Here $g_{ij}$ is the coupling strength in the mean field treatment between
the $i^{th}$ and $j^{th}$ component.
Interestingly, 2 component BECs have been used to create 
vortices \cite{matthews} and study non-equilibrium dynamics \cite{mertes,anderson,nicklas}. 

In this work, we look at the  character of the excitations across the 
quantum phase transition in the 2 component BEC system. 
For a  miscible system, the modes are collective in nature, and the 
condensates move either in-phase or out-of-phase
with each other. For the immiscible system, the excitations are either 
collective or interface excitations.  This work offers a new perspective 
into the nature the miscible-immiscible transition with characterization 
of the excitations. 

There has been previous work on excitation spectra for trapped 
2 component BEC \cite{gordon}. That work studied symmetry breaking as a 
function of particle number and found that a mode goes to zero when the system 
becomes immiscible, but they did not study the nature of the quasiparticles.
Ref. \cite{kim} studied the ground state and
characterized the mode that goes soft (energy goes to zero), here 
we extend the analysis to many low-lying modes of the system.

\section{methods}
To obtain the excitation spectrum, we solve the Bogoliubov de Gennes 
equations \cite{pu,eddy} for a trapped gas with 
contact interactions. First, we must solve the Gross Pitaevskii equation 
for the 2 condensates ($\phi_i$):
\begin{eqnarray}
H_{GP}\phi_i=\left(H_0+\sum_jg_{ij}|\phi_j|^2\right)\phi_i=\mu_i\phi_i.\label{gpe}
\end{eqnarray}
$H_0$ is the kinetic energy and trapping potential and $\mu_i$ is the chemical
potential for the $i^{th}$ component.
We normalize $\phi_i$ so that $\int dV |\phi_i|^2$=1.
Now we consider the ground state 
and its excitations to be of the form:
$e^{-i\mu_i t/\hbar}\{\phi_i+\lambda(u_i e^{-i\omega t}+v_i^* e^{i\omega t})\}$.   
Substituting this into the time dependent version of Eq. (\ref{gpe}) 
and collecting powers of $e^{\pm i\omega t}$ and linear terms in $\lambda$, 
we find the excitations are given by:
\begin{eqnarray}
&&\left(\begin{array}{cc}
H_{GP}-\mu_i&0\\
0&H_{GP}-\mu_i\\
\end{array}\right)
\left(\begin{array}{c}u_i^\alpha\\v_i^\alpha\end{array}\right)\label{bdg}
\\\nonumber&&+\sum_{j}\left(\begin{array}{cc}
g_{ij}\phi_i\phi_j&g_{ij}\phi_i\phi_j\\
g_{ij}\phi_i\phi_j&g_{ij}\phi_i\phi_j
\end{array}\right)
\left(\begin{array}{c}u_j^\alpha\\v_j^\alpha\end{array}\right)
=\omega_\alpha
\left(\begin{array}{c}u_i^\alpha\\-v_i^\alpha\end{array}\right).
\end{eqnarray}
We have assumed $\phi_i$ is real, and that the  different components
have equal number and mass. The second term contains
both the exchange and anomalous term, which couples $u_i^\alpha$ to $u_j^\alpha$ and 
$u_i^\alpha$ to $v_j^\alpha$, respectively. These terms would be non-local if the interaction
was finite range. These excitations are normalized in the
standard way: $\int dV |u^\alpha|^2$-$|v^\alpha|^2=1$ \cite{fetter}.

To perform this work, we focus on the quasi2D case where high resolution 
experimental imaging is possible \cite{chicago,bakr}.
For clarity, we will focus on two examples: one is miscible and 
the other immiscible, both far away from the transition so the character of the
excitations is clear.
We consider $g_{11}=g=$100 and $g_{22}$=1.01g;
for the miscible example, we pick $g_{12}$=0.5g and an immiscible 
example, we pick $g_{12}$=2g, where 
$g=N\sqrt{8\pi} \hbar^2a_s/m l_z$ is the strength of the contact 
interaction, $a_s$ is the 3D s-wave the scattering length 
($a_s\ll l_z$), $l_z=\sqrt{\hbar/m \omega_z }$ is the axial harmonic 
oscillator length and $\omega_z$ is the trapping frequency in the
tightly confined direction.  We only consider $N=N_1=N_2$ and equal masses.
We rescale the equations into trap units, so the energy scale is 
$\hbar\omega_\rho$ and the length scale is $l_\rho=\sqrt{\hbar/m \omega_\rho }$
where $\omega_\rho$ is the trapping frequency in the x-y plane.
We can loosely relate this to experiments, for g=100 if we pick $N$=1000, 
$\omega_z$/$\omega_\rho$=100, and $a_s$=100 $a_0$, 
then this example corresponds to radial trapping frequencies, 
of $2\pi\times$38 Hz for K and $2\pi\times$11 Hz for Cs.
It is worth mentioning that the chemical potentials for each component are about equal
($\mu=\mu_1\sim\mu_2$). More importantly, for the immiscible system, $\mu$ is  
$8.4\hbar\omega_\rho$ and this gives a healing length of $\xi\sim0.35l_\rho$
(for the miscible system $\mu\sim7.4\hbar\omega_\rho$).
Further details of how we solve these equations (\ref{gpe},\ref{bdg}) 
appear in Ref. \cite{CThfb}.

\begin{figure}
\includegraphics[width=0.95\columnwidth]{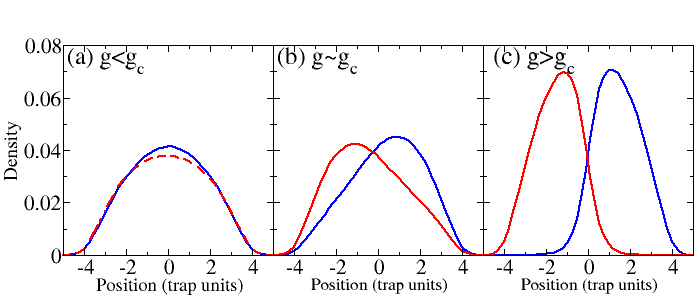}%ground_change 
\caption{(color online) The evolution of the ground state from
miscible (a) to immiscible (c). 
In (a) we show $g_{12}$ only slightly less than $g_{12}^c$, and 
in (b) we show the quantum gas just after
quantum phase transition when the ground state breaks rotational symmetry
and $g_{12}$ only slightly greater than $g_{12}^c$. For
(c) we have shown the standard example of an immiscible gas with $g_{12}/g=2$.
The BEC labeled 1 (2) is shown as blue (red).}
\label{ground}
\end{figure}

\section{results}
In Fig. \ref{ground} we see the ground state changes character as we vary 
$g_{12}$. The component with the smaller $g_{ii}$ is 
more dense in the middle of the trap.
We define $g_{12}^c$ as the value of $g_{12}$ when the ground state
changes character to a broken symmetry state which begins
the immiscible regime.  
When  $g_{12}<g_{12}^c$, the ground state of the system has azimuthal symmetry 
and the 2 BECs overlap, (a). However as $g_{12}$ is increased 
to $g_{12}^c$, the ground state suddenly changes, and the  azimuthal symmetry
is broken, see (b). 
As $g_{12}$ is further increased, the 2 BECs separate further and
decrease their overlap as it becomes energetically costly, (c).
A similar evolution of the ground state as a function of $g_{12}$ was reported 
in Ref. \cite{navarro}.

It is challenging to find the ground state for all $g_{12}$. To do so, 
we use the conjugate gradient method.  We have found
the best initial guess is one with a slightly broken symmetry and poor
overlap with the final group state.  
We seed the noise so that the interface would be along the y axis.
If the overlap between the initial guess and the ground state is too large
then it is easy to get stuck in a local energy minimum. When the 
conjugate gradient method fails to find the true ground state, 
the solutions to the Bogoliubov de Gennes equations have complex eigenvalues. 
When we vary $g_{12}$, the previous solution is thrown out.
In this way we reliably find the 
excitation spectrum of the 2 component BEC with only real eigenvalues.

\begin{figure}
\includegraphics[width=.9\columnwidth]{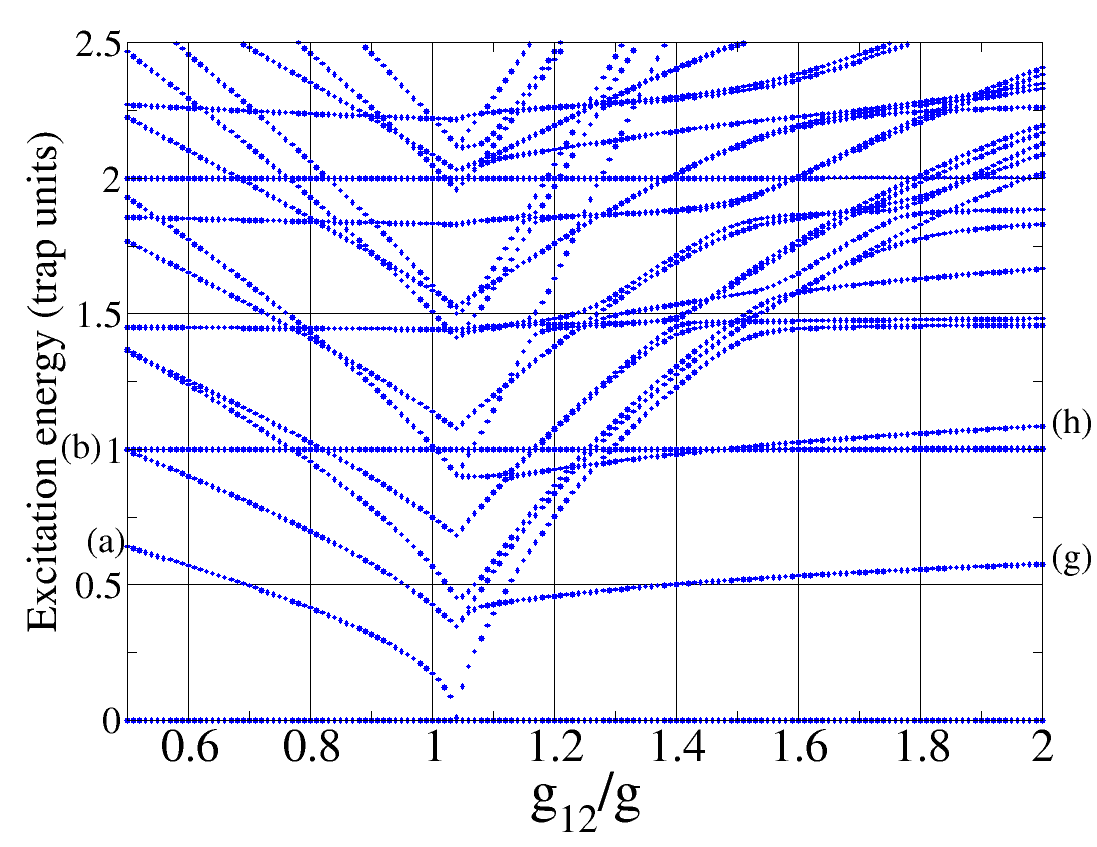} %energytrans
\caption{(color online) Excitation spectrum for a transition from 
miscible to immiscible. The critical $g_{12}^c$ is at 1.04g for trapped example
with $g$=100.
Note the presence of strong avoided crossings and broken degeneracies
after rotational symmetry has been broken, $g_{12}>g_{12}^c$.
For the miscible system as $g$ is increased, the out-of-phase collective excitations
dramatically lower in energy.
Some of the excitations shown in Fig. \ref{quasi} (\ref{quasi2}) are labeled 
on the left (right) of this figure.
}\label{energytrans}
\end{figure}

In Fig. \ref{energytrans} we show Bogoliubov excitation energies 
($\omega_\alpha$) as a 
function of $g_{12}$.  This shows the transition from miscible
to immiscible at $g_{12}^c/g\sim1.04$, where a mode goes soft.
Homogeneous theory predicts this transition at $g_{12}^c=\sqrt{g_{11}g_{22}}$ \cite{pu,eddy}.
The discrepancy is explained by the trap and finite size 
of the gas \cite{ho}. 
In fact, if we were to increase g (keep $g_{ij}/g$ fixed); $g^c_{12}$ 
decreases toward 1. For example if g=400 ($\mu\sim16\hbar\omega_\rho$) 
then $g^c_{12}\sim1.01g$.  
A recent study explored how the trapping and the kinetic energy contributions
impact the criteria for immisciblity \cite{wen}.  Our findings are consistent 
with their results.

To further understand Fig. \ref{energytrans}, we start with $g_{12}<g_{12}^c$
where the systems is miscible. The quasi-particles are readily classified 
based on their azimuthal symmetry and the relative 
motion of the 2 condensates. 
As $g_{12}$ increases towards $g_{12}^c$, many modes decrease in energy.  
Then at $g_{12}^c$ two modes go soft or their energies go to zero.
For $g_{12}>g_{12}^c$, many degeneracies are broken, 
and there are many avoided crossing as $g_{12}$ is further increased.  
This is where the excitations mix and change character.
For the miscible side of spectrum, there
are energy crossings, but they are 
protected symmetry and do not couple.
To further understand this transition, 
we look at the mode which goes soft at $g_{12}^c$.
%###########################################################################

\begin{figure}
%\vspace{-8pt}
\includegraphics[width=0.9\columnwidth]{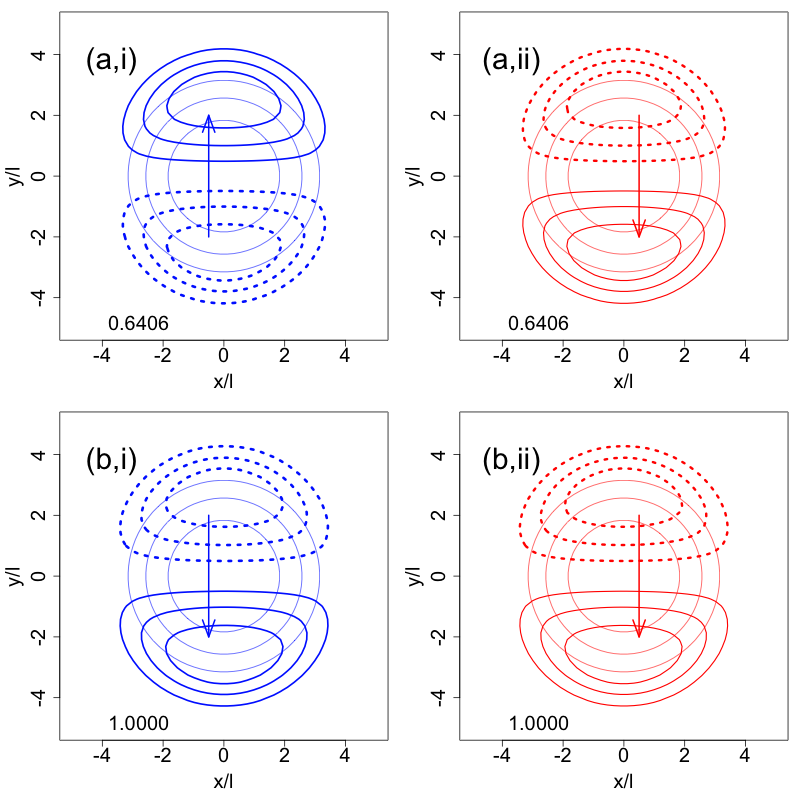}%qp11p5short
\caption{(color online) We show the density perturbations as a function 
of x and y. 
Theses perturbations are co-spatial, but for clarity they have been separated.
The density moves out of the dashed regions and into the 
solid regions. The condensate densities are shown as faint contours. 
In (a) we show the out-of-phase slosh mode where (i) moves up and (ii) moves down.
This mode has no center-of-mass motion. 
In (b) we show the in-phase slosh mode where (i) and (ii) move together.
This is a Kohn mode with center-of-mass motion.  
For this example $g_{11}=g=g_{22}/1.01=2g_{12}$ with $g=100$.
Energy is shown for each quasi-particle in trap units. }\label{quasi}
\end{figure}

In Fig. \ref{quasi} we show density perturbations associated with 
2 low-energy excitations for the miscible system 
(far left of Fig. \ref{energytrans} with $g=100$). 
Density perturbations from Bogoliubov de Gennes theory (T=0) are given by
$\phi_i(u_i^\alpha+v_i^\alpha)$ for the mode $\alpha$.
In Fig. \ref{quasi} we show both
(a) the out-of-phase and (b) the in-phase slosh modes. 
The condensates for this example are co-spatial and nearly identical.
They look similar to those in Fig. \ref{ground} (a). In this figure, we
separated the two components for clarity.
The density moves from the regions define by the 
dashed lines to the regimes defined by the solid lines. 
The color of the perturbations matches the color for the associated condensate. 
We have drawn arrows to illustrate the motion of the density perturbations.
The contours are shown for 0.25, 0.5 and 0.75 of the maximum value of
the perturbation, and condensate density is shown in the background.  
The energy of the mode is reported on the figure in trap units.

In Fig. \ref{quasi} (a) we show a slosh mode, but 
the motions of the 2 BECs are out-of-phase with each other. 
The solid lines coincide with the dashed lines for the other condensate's
motion.
For example, the blue condensate, (i), sloshes from $y<0$ to $y>0$
while the red condensate, (ii), sloshes from $y>0$ to $y<0$.  
There is no center-of-mass motion in this case. In contrast,
(b) shows a mode with center-of-mass motion which is a Kohn mode
of energy 1$\hbar\omega_\rho$.
The motion of each condensate coincides with the other;
both the blue and red condensate slosh from $y>0$ to $y<0$
in phase with each other. 
It is important to note that the out-of-phase slosh mode has the lowest 
energy, and goes soft at the quantum phase transition.

For $g_{12}<g_{12}^c$, modes with $|m|>0$,
where $m$ is the azimuthal quantum number, have a degenerate twin.
In our case with real quasi-particle modes, degenerate modes are related 
by a rotation of $\pi/2m$.
For example in \ref{quasi} (a) and (b) there 
are degenerate twins are just rotated by $\pi/2$.

Referring back to the energy spectra in Fig. \ref{energytrans},
we see that as $g_{12}$ increases towards $g_{12}^c$ the energy of the mode in
Fig \ref{quasi} (a) decreases (while (b) stays at $1\hbar\omega_\rho$).
More generally, all out-of-phase modes significantly lower in energy as
$g$ is increased to $g_{12}^c$.  
In fact, the out-of-phase modes with $m=1$,  $m=2$, $m=3$, and  $m=0$
are all below  $1\hbar\omega_\rho$ at $g_{12}^c$.
Then at $g_{12}^c$ the energy of the out-of-phase slosh
goes to zero and one mode stays zero for $g_{12}>g_{12}^c$.
The other mode (rotated by $\pi/2$) shoots up 
in energy as $g_{12}$ is further increased beyond $g_{12}^c$. 

The ground state spontaneously breaks rotational symmetry - which 
the Hamiltonian has - and this leads to an extra zero energy Goldstone 
mode in the excitations spectrum \cite{gold}. 
There are already two Goldstone modes associated
with broken phase symmetry of each condensate, and they are:
$u_i^{(1)}=v_i^{(1)}=\phi_i$ and $u_i^{(2)}=-v_i^{(2)}=\phi_i$.
If one looks more closely at the third mode with $\omega=0$, 
one finds it is a rotation of the interface  ($u_i^{(3)}=v_i^{(3)}\ne\phi_i$). 
This extra zero energy mode has already been observed in the 
2 component BECs \cite{kim}. 
Goldstone modes have been discussed in more detail for spinor BECs \cite{ho,ueda},
and Ref. \cite{kan} found similar behavior for a Goldstone mode in an attractive
condensate with a Bogoliubov de Gennes treatment.

As we have said for the miscible system, the classification of the modes is
simple: we use azimuthal symmetry and relative motion of the two 
condensates.  But for the immiscible system, there is no rotational symmetry, 
so the characterization of the excitations must be different. 
To study this in more detail, we look at several quasi-particles.

%++++++++++++++++++++++++++++++++++++++++++++++++++++++++++++++++++++++

\begin{figure}
\includegraphics[width=0.98\columnwidth]{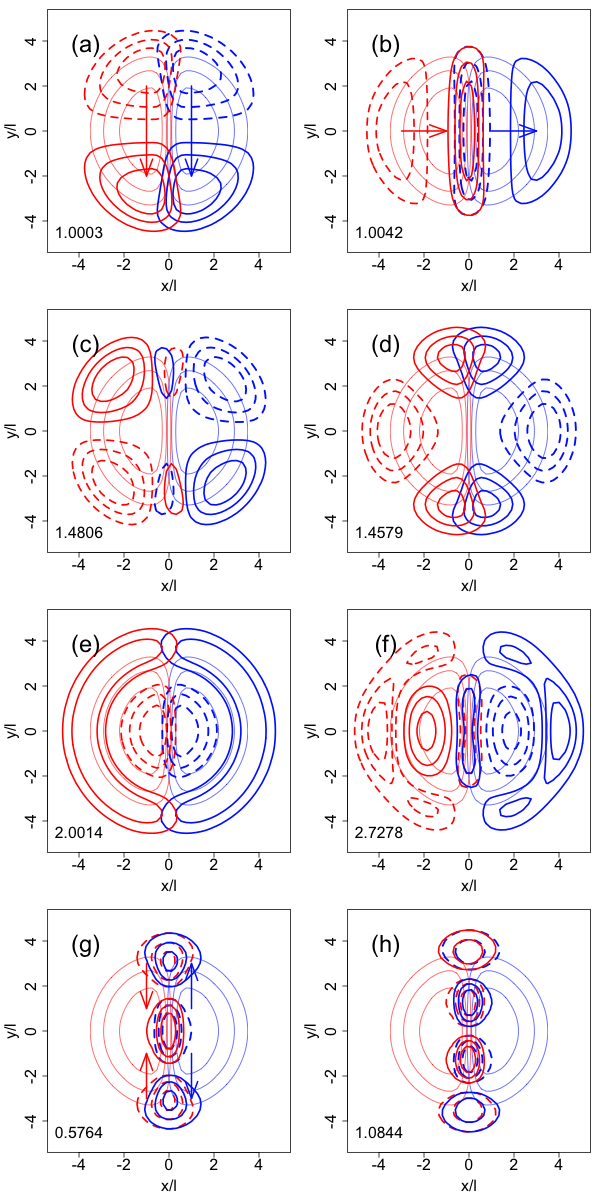} %qp112short
\caption{(color online) The quasi-particle density perturbations for the 
immiscible system, $g_{12}=2$. The x and y axes are in trap units.
(a,b) are slosh modes, (c,d) are quadrupole modes (e,f) are breathing
modes, and (g,h) are interface modes. The energy is reported in trap units
and for this example $g_{11}=g_{22}/1.01=g_{12}/2=100$.}
\label{quasi2}
\end{figure}

In Fig. \ref{quasi2} we show the density perturbations associated 
with quasi-particle excitations for the 
immiscible system ($g_{11}=g$, $g_{22}=1.01g$, $g_{12}=2g$, and g=100).
There are two types of quasi-particle modes:
first, bulk excitations which look like those from a standard condensate, and
second, interface excitations where the excitations are localized to the 
interface between the 2 condensates.  Since there is no azimuthal symmetry, 
to classify 
the bulk modes we need to access how their motion is oriented relative to the 
interface. In Fig. \ref{quasi2}, to depict the motion of the density perturbations, 
we show arrows in 
a few examples.  The density moves from the dashed regions to the 
solid regions. We also show the energy of the excitation in trap units.
The contours are shown for 0.25, 0.5 and 0.75 of the maximum value of
the perturbation and condensate density shown in the background.

The collective modes look like those in a standard BEC,
however the two BECs now act collectively 
to retain the excitation character.
First, we look at the slosh modes of the systems. They are shown in 
Fig. \ref{quasi2} (a,b).  In 
(a) the slosh mode with the center-of-mass displacement parallel to the 
interface (Kohn mode) is shown and in 
(b) a slosh mode and center-of-mass displacement is perpendicular 
to interface (also a Kohn mode) is shown. 
The arrows show that in (a) both the blue and red  condensate sloshes from 
$y>0$ to $y<0$.  For example (b) both the blue and red condensate 
sloshes from $x<0$ to $x>0$.
 
Next we show the quadrupole modes in Fig. \ref{quasi2} (c,d).
(c) shows a quadrupole mode with a nodal line along interface, and
(d) shows a quadrupole mode where the density increases at interface. 
These excitations are typical of a single component BEC
where the excitations are related by a $\pi/2m$ rotation.
But in this case, the 2 BECs collude to make the excitation.  Additionally,
these two excitations are very similar in energy.

We show two breathing modes in Fig. \ref{quasi2} (e,f).
They can be classified as in-phase and out-of-phase motion of the 2 condensates. 
In (e) we show an in-phase breathing mode, where both BECs inhale at once, or 
they both move into or out of the center of the trap in unison.
In (f) we show an out-of-phase breathing mode, where one condensate 
inhales and the other exhales. The energies of these modes are notably
different: 2 and 2.73 $\hbar\omega_\rho$. 

\begin{figure}
\includegraphics[width=0.96\columnwidth]{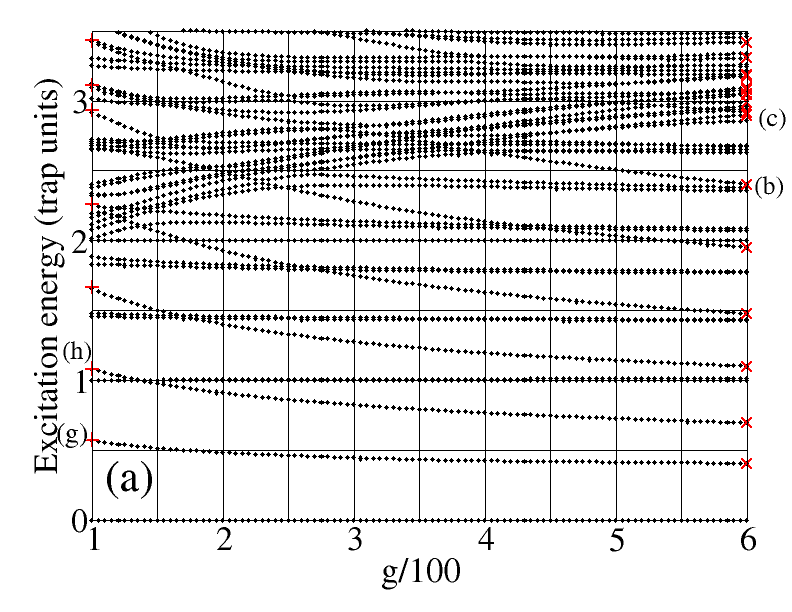}%energyN
\\\centerline{
\includegraphics[width=0.32\columnwidth]{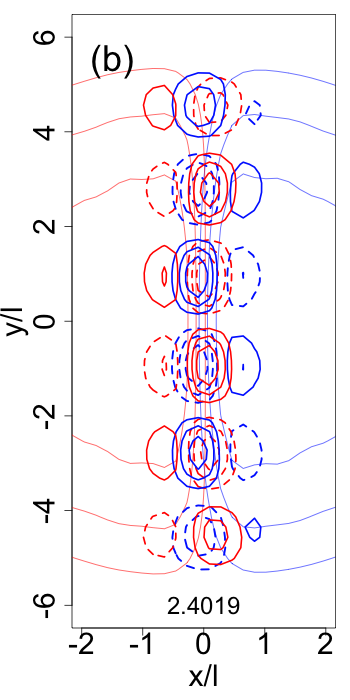}%qpINT1
\includegraphics[width=0.64\columnwidth]{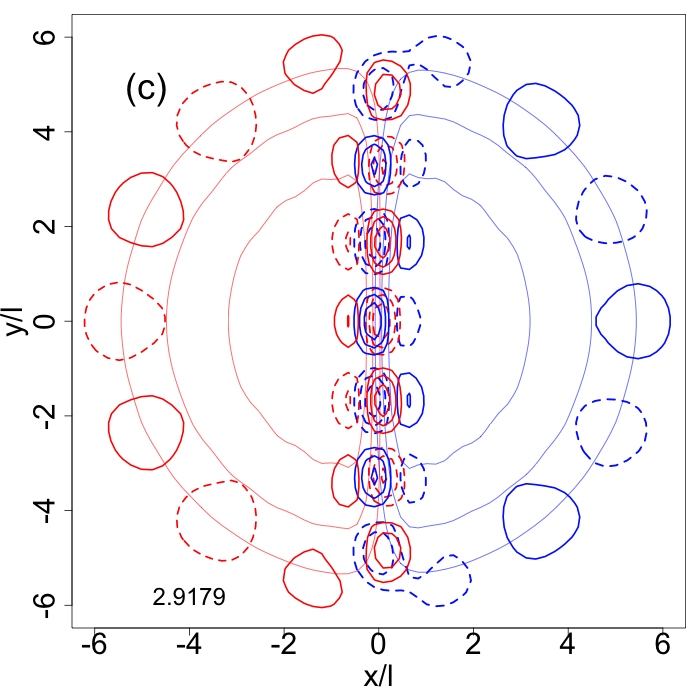}%qpINT2
}
\caption{(color online) The excitation spectrum as a function of g
for the immiscible example ($g_{12}/2=g_{11}=g_{22}/1.01$). 
The quasiparticles from Fig. \ref{quasi2} are on the far left side of spectrum.
The interface excitations are marked by $\times$ or $+$ on either side of 
the figure.
The energies of  the quasiparticle shown in Fig. \ref{quasi2} (g,h) are labeled.
In (b) and (c) the density perturbations for the interface excitations when
$\mu\sim20\hbar\omega_\rho$ or $g$=600, they are labeled on the right side of (a).
Their energy is reported in trap units.}
\label{energyN}
\end{figure}

There is another class of excitation in the immiscible 2 component BEC:
interface excitations. 
Two examples are shown in Fig. \ref{quasi2} (g, h).
These excitations are localized along the interface, and in
general they are out-of-phase excitations, 
i.e. the density of one moves to where the other is leaving.
Note these are low energy excitations, in fact (g) is the lowest energy
excitations, for the system at 0.58 $\hbar\omega_\rho$ and (h) is only 
slightly higher than the two Kohn modes at 1.08 $\hbar\omega_\rho$.
If $g$ is increased the mode in (h) will decrease below 1$\hbar\omega_\rho$.
So if the chemical potential is increased, then the interface modes 
become lower in energy. To illustrate this, we look at how the excitation energies
change as a function of $g$ while keeping the ratio of the interactions
fixed ($g_{12}/2=g_{11}\sim g_{22}$).  This is shown in Fig \ref{energyN} (a).
For reference, on the far right where $g$=600 and $\mu\sim20\hbar\omega_\rho$,
we have marked the interface modes with red $\times$'s, there are 16 interface
modes with energy under 3.5$\hbar\omega_\rho$.
In contrast, on the far left where $g=100$, we have marked the interface modes 
with red $+$'s and there are only 7 modes under 3.5$\hbar\omega_\rho$.
As one moves up in excitation energy, each new excitation simply adds another 
bend to the interface. 
One more important point of Fig. \ref{energyN} (a), is that 
as $g$ increases the energies of interface modes decrease.

Fig. \ref{energyN} (b) and (c) show two examples of the higher energy interface 
modes.
First, Fig. \ref{energyN} (b) is a mode with 6 bends in the interface.
An interesting point about (c) is that it is a hybrid mode, it also has
some density perturbations near the edge of the gas, away from the
interface.  The mode shown in (c) is in a
region where several modes are crossing and their
character is changing.  If we further increase $g$, the interface
mode lowers in energy and  loses it collective nature and
looks more like mode in (b).
Related excitations have been studied in non-equilibrium simulations of 
2 component BECs where Rayleigh-Taylor instabilities have been predicted
 \cite{rt1,rt2,rt3}. In these studies, the value of $g_{11}$ is changed and this
drives one BEC into the other, the interface then becomes unstable and a 
Rayleigh-Taylor instability forms.

\section{conclusions}
In conclusion, we have characterized the excitations of a 2 component BEC 
within the Bogoliubov de Gennes framework. We found that as $g_{12}$ is
increased from a miscible regime to an immiscible regime, 
Fig. \ref{energytrans},  generally, all of the out-of-phase excitations
lower in energy.  The energy of the out-of-phase slosh mode goes to zero, 
Fig. \ref{quasi} (a). 
This mode becomes new Goldstone modes when the rotational symmetry is 
spontaneously broken in the immiscible system.
We looked at the  excitations of the immiscible systems when
$g=g_{12}/2=g_{11}=g_{22}/1.01$  in  Fig. \ref{quasi2}. We found that there are bulk
modes which look similar to the excitations of a single trapped BEC.  
There are also excitations localized at the boundary between the condensates.  
One of these interface modes is lowest energy mode when the BECs are immiscible, 
and there are many other low energy interface modes, 
see the red $+$'s or x's in Fig. \ref{energyN} (a).
Furthermore, if one goes to a more strongly interacting BEC regime 
(while still immiscible), the interface modes lower in energy. 

Future work will seek to understand the relationship between
Bogoliubov excitations across the miscible-immiscible transition 
and critical phenomena.
The effect of temperature on this system will be studied within 
the Hartree Fock Bogoliubov framework, where Bogoliubov excitations
are thermally occupied. Additionally,
we will look at how this collective excitation changes with non-local dipolar 
interactions \cite{CThfb,roton}.

\begin{acknowledgments}
The author gratefully acknowledges support through a
LDRD ECR grant, LANL which is operated by LANS, LLC for the NNSA 
of the U.S. DOE under Contract No. DE-AC52-06NA25396.
This research was supported in part by the National Science Foundation 
under Grant No. NSF PHY11-25915.
The author also gratefully acknowledges conversations 
with E. Timmermans, L. A. Collins, and R. M. Wilson.
\end{acknowledgments}
\bibliographystyle{amsplain}

\end{document}